\documentclass[
 amsmath,amssymb,
 aps,
 twocolumn,
 secnumarabic,
 prb,
]{revtex4-1}

\usepackage{graphicx}% Include figure files
\usepackage{dcolumn}% Align table columns on decimal point
\usepackage{bm}% bold math
\usepackage{color}
\usepackage{stmaryrd}
\newcommand{\bk}{\mathbf{k}}
\newcommand{\bq}{\mathbf{q}}
\newcommand{\bK}{\mathbf{K}}
\newcommand{\bsig}{\mathbf{\sigma}}
\newcommand{\br}{\mathbf{r}}
\newcommand{\dg}{\dagger}
\newcommand{\bd}{\mathbf{d}}
\newcommand{\re}{\mathrm{e}}
\newcommand{\rd}{\mathrm{d}}

\newcommand{\MF}{\mathrm{MF}}
\newcommand{\dr}{\mathrm{d}r}

\newcommand{\cA}{{\cal A}}
\newcommand{\cC}{{\cal C}}
\newcommand{\cT}{{\cal T}}

\newcommand{\cP}{{\cal P}}
\newcommand{\cR}{{\cal R}}
\newcommand{\cH}{{\cal H}}

\begin{document}

%\preprint{APS/123-QED}

\title{Majorana-Hubbard model on the honeycomb lattice}% Force line breaks with \\

\author{Chengshu Li}
\email{chengshu@phas.ubc.ca}
\author{Marcel Franz}
\email{franz@phas.ubc.ca}
\affiliation{Department of Physics and Astronomy \& Quantum Matter Institute, University of British Columbia, Vancouver, British Columbia V6T 1Z1, Canada}

\date{\today}% It is always \today, today,

\begin{abstract}
Phase
diagram of a Hubbard model for Majorana fermions on the honeycomb
lattice is explored using a combination of field theory,
renormalization group and mean-field
arguments, as well as exact numerical diagonalization. Unlike the previously studied versions of the model we find
that even weak interactions break symmetries and lead to interesting
topological phases.  We establish two
topologically nontrivial phases  at weak
coupling, one gapped with chiral edge
modes and the other gapless with antichiral edge modes. At strong coupling a mapping onto a novel frustrated
spin-${1\over 2}$ model suggests a highly entangled spin liquid ground state. 

\end{abstract}
\maketitle

%\tableofcontents

\section{Introduction}
The Hubbard model has long served as a platform for explorations of
strongly interacting systems
[\onlinecite{Hirsch1985,White1989,Bickers1989,Sorella1992}]. It has been
extensively studied for spinful and spinless fermions and bosons, in
different dimensions, and on various lattices, serving as a rich
source of new physics, and providing insights into phenomena ranging from
metal-insulator transition to high-temperature superconductivity. 

In recent years,  theoretical
[\onlinecite{Read2000,Kitaev2001,Stern2008,Nayak2008,Fu2008,Lutchyn2010,Oreg2010,Alicea2012,Beenakker2013,Elliott2015}]
and experimental
[\onlinecite{Mourik2012,Das2012,Deng2012,Rokhinson2012,Finck2013,Hart2014,Nadj-Perge2014,Kammhuber2017,He2017}]
developments revealed Majorana fermions [\onlinecite{Majorana1937}],
as readily observable emergent particles in certain condensed matter
systems. This motivates a thorough study of their physical properties
in various situations. Specifically, Majorana-Hubbard models have been
formulated to explore the effects of interactions between localized
Majorana zero modes. A one-dimensional (1D) Majorana-Hubbard model was extensively
studied using combined techniques of field theory, renormalization group and density matrix renormalization group
 with a rich phase diagram and a supersymmetric phase transition
identified [\onlinecite{Rahmani2015b,Rahmani2015a,Cobanera2015,Sannomiya2017,Fendley2018}]. Similar phase transitions were
discovered in a ladder model [\onlinecite{Zhu2016}] and on a 2D square
lattice [\onlinecite{Affleck2017}]. Reference\ \onlinecite{Chiu2015} further argued that these models may be relevant to
Majorana zero modes localized near vortices in the Fu-Kane
superconductor [\onlinecite{Fu2008}], realized at the proximitized surface of
a 3D topological insulator and recently confirmed in a series of
experiments [\onlinecite{Jia2016a,Jia2016b}]. 

\begin{figure}
\centering
\includegraphics[width=7cm]{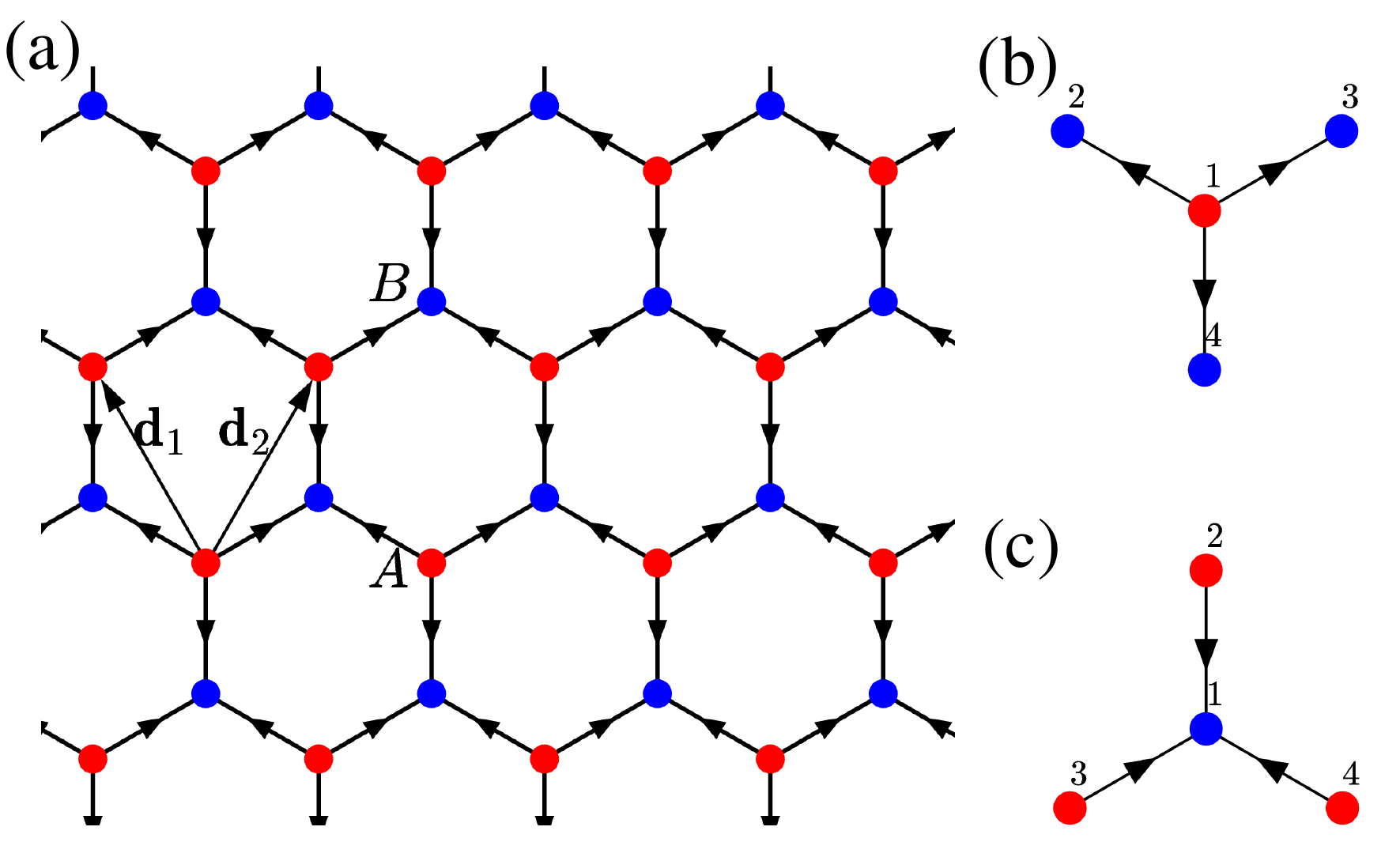}
\caption{(a) Lattice structure of the model. The arrow indicates the sign of the hopping terms. (b,c) The ordering of the four operators in $g_1$ and $g_2$ terms, respectively.}\label{lattice}
\end{figure}

In this paper we report on a comprehensive study of the
Majorana-Hubbard  model on the honeycomb lattice. 
The honeycomb lattice has been of interest to both theoretical
[\onlinecite{Semenoff1984,Haldane1988,Kane2005}]  and experimental
[\onlinecite{Geim2007,Neto_RMP,Cao2018}] communities due to its simplicity and
its remarkable wealth of physical
properties. The model Hamiltonian we explore here reads $\cH=\cH_0+\cH_{\mathrm{int}}$ with 
\begin{equation}\label{h0}
\begin{split}
\cH_0&=it\sum_{\langle ij\rangle}\eta_{ij}\alpha_i\beta_j,\\
\cH_{\mathrm{int}}&=g_1\sum_{\Ydown}\alpha_i\beta_j\beta_k\beta_l
+g_2\sum_{\Yup}\beta_i\alpha_j\alpha_k\alpha_l.
\end{split}
\end{equation}
The Majorana operators on the $A$ $(B)$ sublattice, $\alpha_i$
$(\beta_i)$, obey
$\alpha_i^\dagger=\alpha_i$, $\beta_i^\dagger=\beta_i$ and 
\begin{equation}
\{\alpha_i,\alpha_j\}=\{\beta_i,\beta_j\}=2\delta_{ij}, \ \  \ \{\alpha_i,\beta_j\}=0.\label{anti-comm}
\end{equation}
The hopping amplitude $t>0$ sets the energy scale and the phase factors
$\eta_{ij}=\pm 1$ are constrained by the Grosfeld-Stern rule
[\onlinecite{Grosfeld2006}]. We choose a gauge as in Fig. \ref{lattice}(a) to
minimize the unit cell. Figures \ref{lattice}(b) and \ref{lattice}(c) show the order of
the Majorana operators in the two interaction terms, representing the
most local interactions possible.

\begin{figure}
\centering
\includegraphics[width=8.7cm]{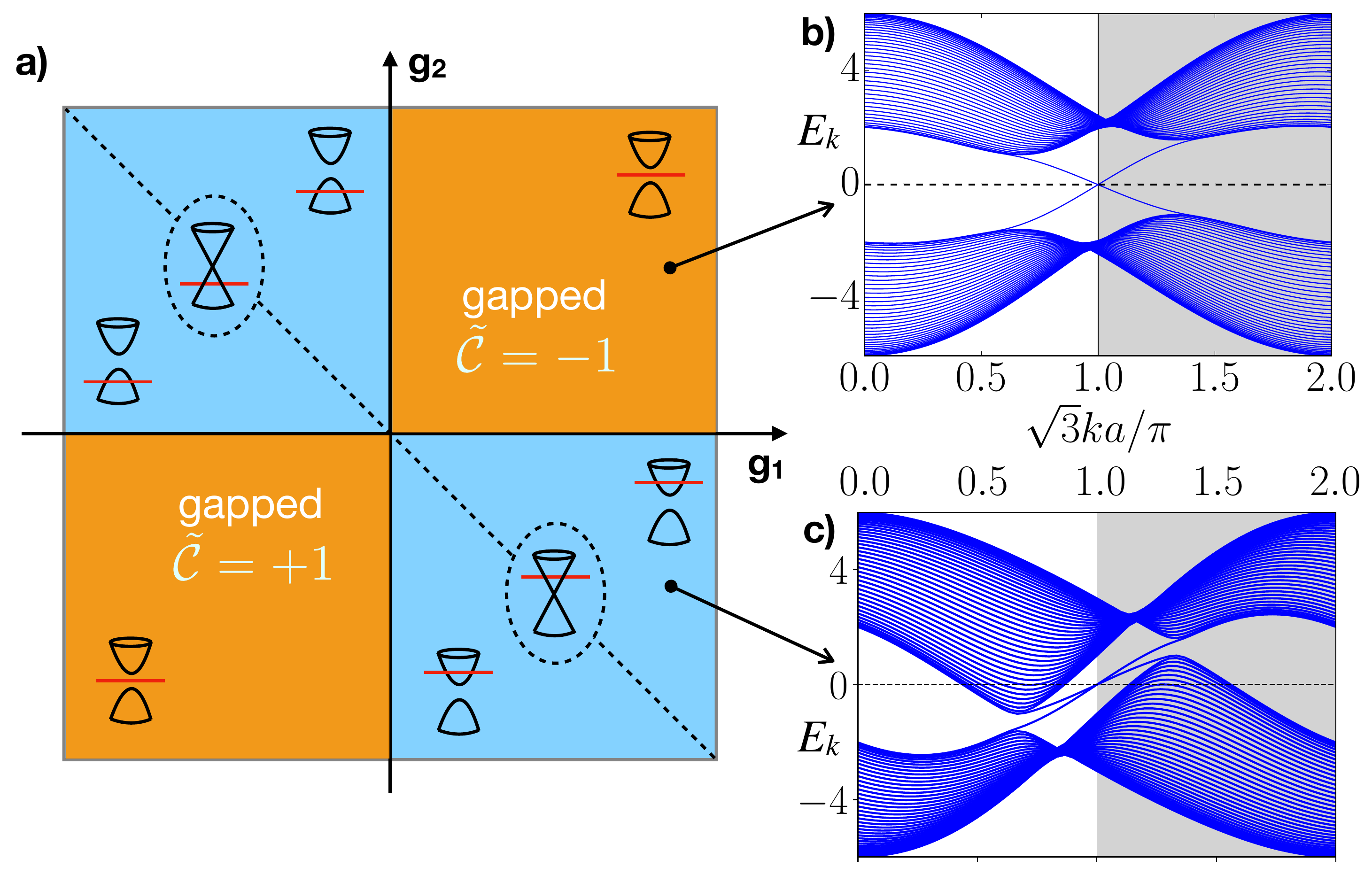}
\caption{(a) Phase diagram of the interacting Majorana model at
  weak coupling. MF Hamiltonian Eq.\ (\ref{hmf}) on a strip 
with the zig-zag boundary (with $\tau_1=\pm0.1,\ \tau_2=-0.15$) is
used to illustrate  the energy spectra 
 in the gapped Majorana Chern insulator phase (b) and the gapless
  Majorana metal phase (c).}\label{fig-3}
\end{figure}

The noninteracting model with $g_1=g_2=0$ exhibits a unique
ground state with linearly dispersing single particle
excitations near the $\pm\bK$ corners of the
hexagonal Brillouin zone, analogous to graphene
[\onlinecite{Neto_RMP}]. Unlike in graphene and in the previously studied
Majorana-Hubbard models
[\onlinecite{Raghu2008,Rahmani2015b,Rahmani2015a,Cobanera2015,Fendley2018,Zhu2016,Affleck2017}], where weak
interactions initially do not change the nature of such a state, we find dramatic
effects induced by $\cH_{\mathrm{int}}$ that occur already at
infinitesimal coupling strength. Our results are summarized in Fig.\
\ref{fig-3}(a) which shows the phase diagram of the interacting model at
weak to intermediate coupling. Except for the line $g_2=-g_1$ 
interactions give rise to a gap in the excitation spectrum. As
explained below in the two quadrants with $g_1g_2>0$ the system can be
characterized as a Majorana Chern insulator with Chern number
$\tilde{\cC}=-{\rm sgn}(g_1)$ and topologically
protected chiral edge modes, Fig.\ \ref{fig-3}(b). This phase belongs to the class D of the Altland-Zirnbauer classification [\onlinecite{Chiu2016}]. In the other two
quadrants one obtains a Majorana metal with topologically protected
antichiral edge modes [\onlinecite{Colomes2018}] illustrated in  Fig.\
\ref{fig-3}(c). At stronger coupling ($g_1,g_2\gtrsim 5t$)  our exact
diagonalization (ED) numerics suggests a transition to a strongly
entangled gapped phase with a doubly degenerate ground state.   

\section{Symmetries}
In addition to discrete translations and
${2\pi\over 3}$ rotations the non-interacting model is invariant
under inversion $\cP$ and two reflections $\cR_1$ and $\cR_2$
indicated in Fig.\ \ref{fig-2}(a). As explained in Appendix A 
some of these operations must be supplemented by an
appropriate Z$_2$ gauge transformation, which we henceforth denote by $\cA$,
in order to become symmetries.  It is easy to deduce that
$\cH_{\mathrm{int}}$ respects $\cP\cA$ if $g_1=g_2$ and respects
$\cR_2\cA$ if $g_1=-g_2$. Under $\cR_1$ the Hamiltonian
$\cH(t,g_1,g_2)$ maps onto $\cH(t,-g_1,-g_2)$.
In addition $\cH_0$
is invariant under antiunitary time-reversal symmetry $\tilde{\cT}$
which maps  $(\alpha_j,\beta_j)\to
(\alpha_j,-\beta_j)$ and $i\to -i$. However, $\cH_{\rm int}$ breaks
$\tilde{\cT}$ for any  nonzero $g_1$, $g_2$.

If the Majorana fermions are realized in vortices of the Fu-Kane
superconductor then, based on the above analysis, we have $g_1=g_2$ if
the lattice is composed of vortices only, but $g_1=-g_2$
if sublattice A has vortices and sublattice B antivortices (or vice
versa). This is because both $\cP$ and $\cR_2$ interchange the
sublattices and inversion preserves vorticity while reflection maps
vortex onto an antivortex. If the honeycomb lattice becomes distorted
such that it no longer respects $\cP$ and $\cR_2$,
then in general there will be no constraint on $g_1$ and $g_2$. In the
following we analyze the model for arbitrary coupling constants but
pay particular attention to the high-symmetry cases discussed above.

\section{Low-energy theory}
It is instructive to
examine the low-energy effective theory constructed by expanding the Majorana
fields around the two nodal points at $\pm\bK$ (see Appendix B for details).  One thus obtains
\begin{equation}\label{hc0}
\cH_0\simeq -v\int \rd^2r\sum_{\sigma=\pm}\sigma\left(\alpha_{
    \sigma}\partial_\sigma\beta_{\bar\sigma}+ \beta_{\bar\sigma}\partial_\sigma\alpha_{
    \sigma}\right),
\end{equation}
where $(\alpha_\pm,\beta_\pm)$
are the long-wavelength components of the Majorana fields near points
$\pm\bK$, ${\bar \sigma}=-\sigma$, $v=3ta$ is the
characteristic velocity, $a$ denotes the lattice constant, and
$\partial_\pm=(\partial_x\pm i\partial_y)$.  Similarly we find
\begin{eqnarray}\label{hc1}
\cH_{\rm int}\simeq {24\sqrt{3}}a\int
d^2r\sum_{\sigma=\pm}\bigl[g_1&&\beta_{
    \sigma}\beta_{\bar\sigma}(\alpha_{
                                \sigma}\partial_{\sigma}\beta_{\bar\sigma}) \\
  &&+ g_2 \alpha_{\bar\sigma} \alpha_{\sigma} (\beta_{\bar\sigma}\partial_\sigma\alpha_{
    \sigma})\bigr]. \nonumber
\end{eqnarray}
\begin{figure}[t]
\centering
\includegraphics[width=6cm]{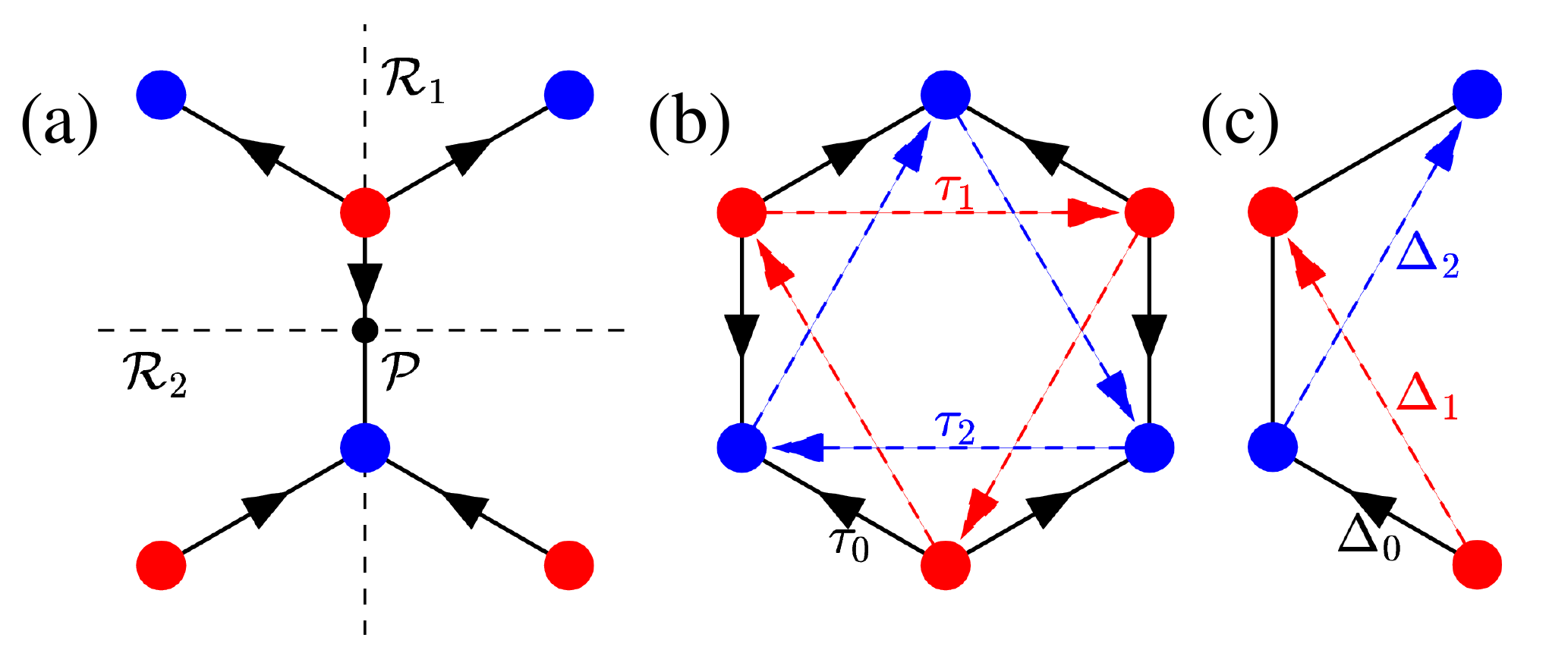}
\caption{(a) Reflection symmetry $\mathcal{R}_{1,2}$ and inversion symmetry $\mathcal{P}$. (b) The sign convention of the MF Hamiltonian. (c) The definition of the order parameters $\Delta_a$.}\label{fig-2}
\end{figure}
As in Ref.\ \onlinecite{Affleck2017} standard renormalization group scaling
arguments indicate that interactions are irrelevant in the low-energy
theory. The Majorana fields have scaling dimension 1 which gives
$\cH_{\rm int}$ dimension 5. The marginal dimension in (2+1)D theory
is 3 so the interactions are strongly irrelevant. Naively, one would
thus expect the system to remain gapless for weak interactions.
 We find, however, that this is not the case for the problem
at hand due to the special structure of the interaction Hamiltonian
(\ref{hc1}). We notice that terms in brackets in Eq.\ (\ref{hc1}) coincide with
those forming the kinetic part $\cH_0$.  Clearly terms present in
$\cH_0$ must have a nonzero vacuum expectation value $\langle\alpha_{
  \sigma}\partial_\sigma\beta_{\bar\sigma}\rangle\neq 0$ and
$\langle\beta_{\bar\sigma}\partial_\sigma\alpha_{\sigma}\rangle\neq
  0$, and it is easy to see that these expectation values will act as
  mass terms when inserted  into $\cH_{\rm int}$. 

If we denote  the above expectation values by $m$  then by symmetry we
expect $\langle\alpha_{
  \sigma}\partial_\sigma\beta_{\bar\sigma}\rangle
=\langle\beta_{\bar\sigma}\partial_\sigma\alpha_{ \sigma}\rangle=\sigma
m$. Replacing the relevant terms in $\cH_{\rm int}$ by
their expectation values and neglecting fluctuations the full
low-energy 
Hamiltonian becomes 
\begin{equation}\label{hc2}
\cH\approx-\int \rd ^2r\sum_{\sigma=\pm} \sigma\Psi^\dagger_\sigma
\begin{pmatrix}
-g_2M & -v\partial_{\bar\sigma} \\
v\partial_{\sigma} & g_1M 
\end{pmatrix}
\Psi_\sigma,
\end{equation}
where $\Psi_\sigma=(\alpha_\sigma,\beta_\sigma)^T$ and $M=24\sqrt{3}am$. Assuming translation invariance the spectrum of $\cH$
is easily obtained by passing to the momentum representation,
\begin{equation}\label{hc3}
E_{k,\sigma}={1\over 2}\sigma M(g_1-g_2)\pm\sigma\sqrt{v^2k^2+{1\over 4}M^2(g_1+g_2)^2},
\end{equation}
where $k=(k_x^2+k_y^2)^{1/2}$.
We observe that interactions produce a gap in the Majorana
excitation spectrum except when $g_2=-g_1$. 
In addition, unequal interaction strengths $g_1\neq g_2$
cause an offset in energy between  two inequivalent nodal
points $\pm\bK$. These considerations lead to the weak-coupling phase diagram
outlined in Fig.\ \ref{fig-3}(a).

When interpreting Eq.\ (\ref{hc3}) one must keep in mind that Majorana
fermions carry half the degrees of freedom of ordinary complex
fermions. Because of this only half of the states implied by
$E_{k,\sigma}$ are physical. Customarily one can either focus on
positive-energy states at all momenta or, equivalently take all
energies but restrict to one-half of the Brillouin zone. In
illustrating various phases of the model we take the latter point of
view and focus on states near $+\bK$.

\section{Mean-field theory}
The low-energy analysis suggests that at
weak coupling accurate results can be obtained using mean-field (MF)
decoupling of the lattice interaction terms Eq.\ (\ref{h0}). As an
example we may approximate $\alpha_i\beta_j\beta_k\beta_l\to
\langle\alpha_i\beta_j\rangle\beta_k\beta_l$ where the expectation
value lives on the nearest neighbor bond (terms already present in
$\cH_0$) and the operator product  $\beta_k\beta_l$ describes coupling between next
nearest neighbors. 
This motivates study of the MF Hamiltonian with first and
second neighbor hoppings
\begin{equation} \label{hmf}
\cH_{\MF}=i\sum_{\langle ij\rangle}\eta_{ij}\tau_0\alpha_{i}\beta_{j}+i\sum_{\langle\langle ij\rangle\rangle}\eta_{ij} (\tau_1\alpha_{i}\alpha_{j}+\tau_2\beta_{i}\beta_{j}),
\end{equation}
with the signs specified in Fig. \ref{fig-2}(b). We expect the
eigenstates of  $\cH_{\MF} $ to capture the essential physics of the problem at weak
coupling and in the following we employ them as variational
wave functions for the full Hamiltonian (\ref{h0}), parametrized by
$\{\tau_a\}$. 
\begin{figure}
\centering
\includegraphics[scale=0.3]{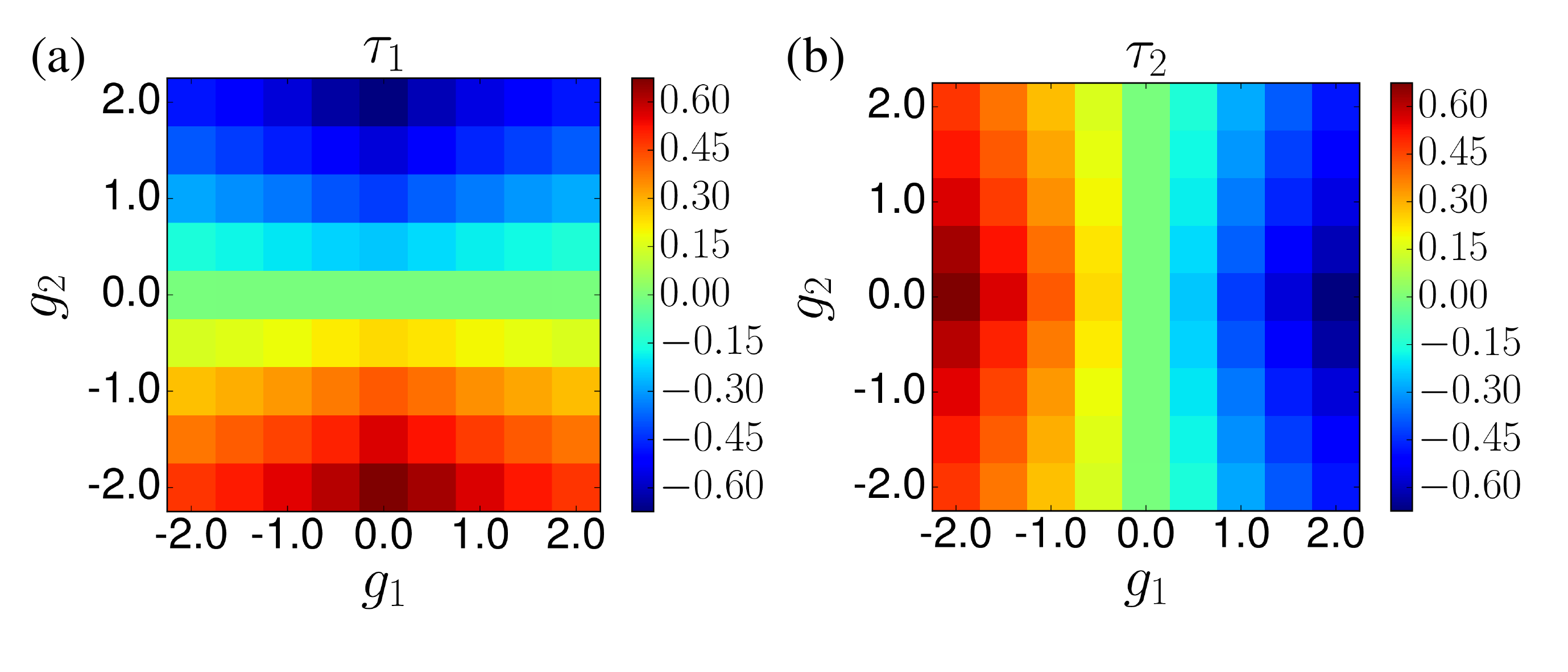}
\caption{Self-consistent solution of the MF equations (\ref{sc}).}\label{fig-4}
\end{figure}

Before linking $\cH_{\MF}$ to the interacting Hamiltonian, it is
useful to understand its properties. In $k$ space, the Hamiltonian
reads
\begin{equation}
\cH_{\MF}=\sum_{\bk}\Psi_{\bk}^{\dg}
\begin{pmatrix}
4\tau_1D_2(\bk) & 2\tau_0D_1(\bk)\\
2\tau_0D_1^*(\bk) & -4\tau_2D_2(\bk)
\end{pmatrix}
\Psi_\bk+E_0,
\end{equation}
where $\Psi_\bk=(\alpha_\bk,\beta_\bk)^T$, 
\begin{equation}
\begin{split}
D_1(\bk)&=i(1+\re^{i\bk\cdot\bd_1}+\re^{i\bk\cdot\bd_2})\\
D_2(\bk)&=-\sin\bd_1\cdot\bk+\sin\bd_2\cdot\bk+\sin(\bd_1-\bd_2)\cdot\bk
\end{split}
\end{equation}
and $E_0=2(\tau_2-\tau_1)\sum_{\bk}D_2(\bk)$. The MF spectrum is
\begin{equation}\label{E_pm}
E_{\bk,\pm}=2(\tau_1-\tau_2)D_2\pm 2\epsilon_{\bk},
\end{equation}
with $\epsilon_{\bk}=\sqrt{\tau_0^2|D_1|^2+(\tau_1+\tau_2)^2D_2^2}$.

As usual, the fact that $\alpha_{-\bk}=\alpha_{\bk}^\dg$ and
$\beta_{-\bk}=\beta_{\bk}^\dg$ introduces a redundancy in the
$k$ space, and we restrict ourselves to half of the BZ. We also take
$\tau_0=1$ without loss of generality. The phase diagram of the model
is then the same as in Fig. \ref{fig-3}(a) with $(g_1,g_2)$ replaced
with $(-\tau_2,-\tau_1)$. 
\begin{figure*}
\centering
\includegraphics[width=12.5cm]{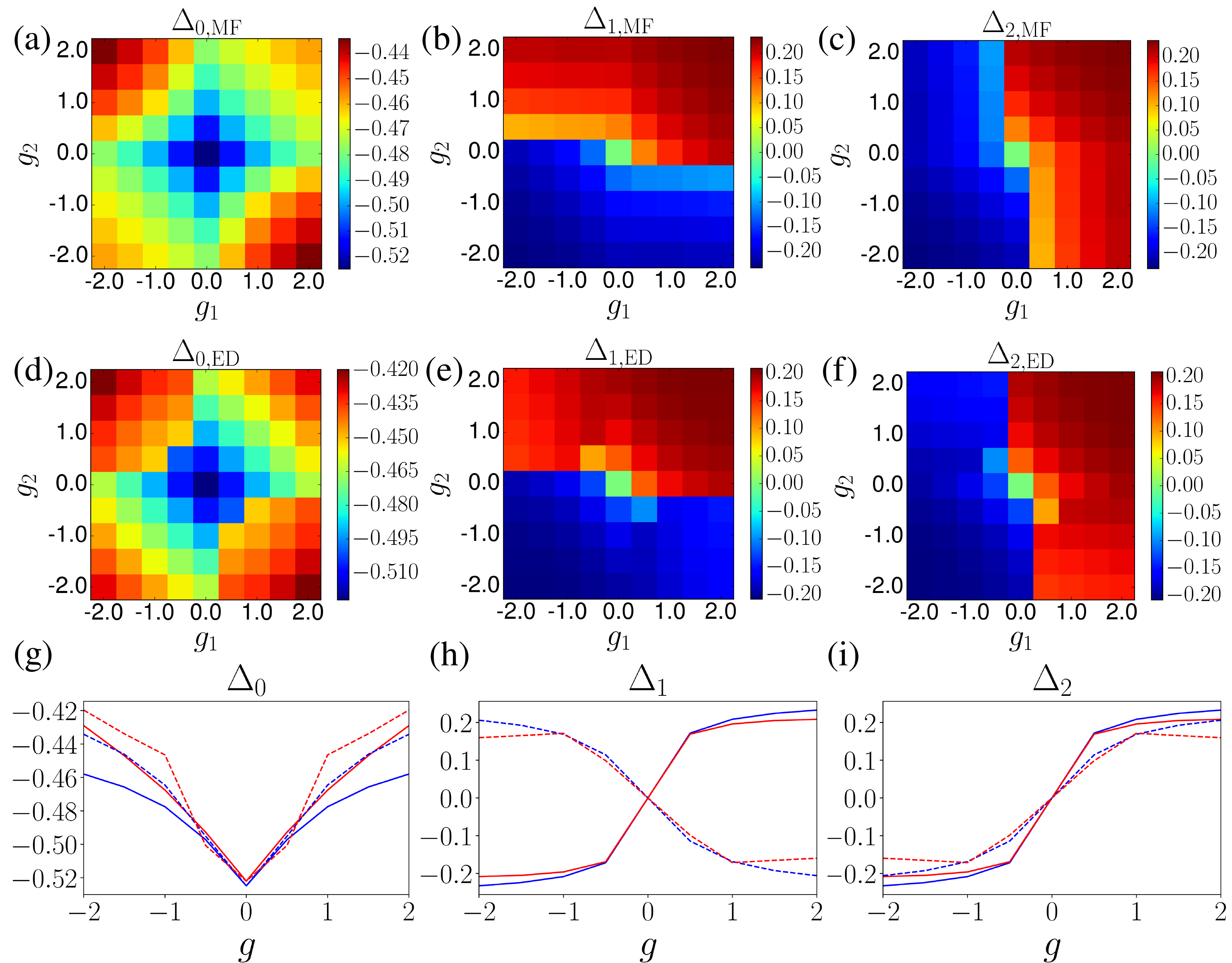}
\caption{Order parameters $\Delta_a$ (a)-(c) from MF calculations
  and (d)-(f) from ED using a cluster of 32 lattice sites with periodic
  boundary conditions. (g)-(i) Quantitative comparison between MF (blue) and ED (red) results, with solid line $g_1=g_2=g$ and dash line $g_1=-g_2=g$.}\label{fig-5}
\end{figure*}

The MF Hamiltonian above resembles the Haldane model
[\onlinecite{Haldane1988}]; thus we expect topologically protected edge modes
in a system with boundaries. In fact, we can readily calculate the Chern number
from the bulk solutions with the caveat that the redundancy of the
$k$-space Hamiltonian implies a Majorana edge mode. For the insulating
phases, $\tau_1\tau_2>0$, the Chern number is 
\begin{equation}
\widetilde{\mathcal{C}}=\mathrm{sgn}(\tau_1)=\mathrm{sgn}(\tau_2).
\end{equation}
We calculate numerically the energy spectrum of the system placed on a
strip  with a zig-zag boundary along the $x$ direction. The energy spectra
for $\tau_1=\pm0.1,$ and  $\tau_2=-0.15$ are shown in
Figs. \ref{fig-3}(b) and \ref{fig-3}(c). The edge modes are clearly present in both
cases. They are chiral for $\tau_1\tau_2>0$ and antichiral for
$\tau_1\tau_2<0$. Chiral edge modes propagate in the opposite
direction on two opposite edges and are protected by the bulk
invariant  $\tilde{\cal C}$. Antichiral edge modes propagate in the same
direction and are protected, to a lesser degree, by their real-space
segregation from the bulk modes [\onlinecite{Colomes2018}].

Now we use the ground state $|\Psi_{\MF}\rangle$ of the MF Hamiltonian
as a variational ansatz to analyze the interacting problem. As
outlined in Appendix C, the requirement that
$\langle\Psi_{\MF}|\cH|\Psi_{\MF}\rangle$ is minimized gives the
MF equations for parameters $\tau_a$,
\begin{equation}\label{sc}
\begin{split}
\tau_0&=t+g_1\Delta_2+g_2\Delta_1,\\
\tau_1&=g_2\Delta_0,\\
\tau_2&=g_1\Delta_0,
\end{split}
\end{equation}
where the order parameters $\Delta_0=i\langle\alpha_1\beta_2\rangle$,
$\Delta_1=i\langle\alpha_1\alpha_2\rangle$, and
$\Delta_2=i\langle\beta_1\beta_2\rangle$ are defined on bonds
specified in Fig. \ref{fig-2}(c). The expectation values are taken with
respect to $|\Psi_{\MF}\rangle$ and  are functions of the variational
parameters $\{\tau_a\}$. They can be expressed as momentum space sums
involving $D_1(\bk)$,  $D_2(\bk)$, and $\epsilon_\bk$,
which we give in Appendix C.

We solve the MF equations (\ref{sc}) by numerical iteration, and the
results are  summarized in Figs.\ \ref{fig-4} and \ref{fig-5}. We
find that $\Delta_0\simeq -0.5 t$ when $g_1=g_2=0$ and  changes very
little with interactions  [Fig.\ \ref{fig-5}(a]. Equations\ (\ref{sc}) then imply that $\tau_1$
and $\tau_2$ become nonzero for arbitrarily weak interaction
strengths $g_{1,2}$.
As a result, we see that an effective second neighbor hopping is
introduced by an infinitesimal interaction strength whereby the system
becomes gapped. MF theory is therefore in full agreement with our
field-theoretic low-energy analysis.

The above instability of the gapless phase is to be contrasted with
the results on the square lattice  [\onlinecite{Affleck2017}], where
interaction strength $|g|\simeq 0.9 t$ is required for the system to
enter a gapped phase. This contrasting behavior can be understood from
symmetry considerations. As in graphene the gapless spectrum near Dirac
points is protected here by a combination of inversion $\cP$ and time
reversal ${\tilde \cT}$. While the full interacting Hamiltonian in
Ref.\  \onlinecite{Affleck2017} respects these symmetries, ${\tilde \cT}$ is
explicitly broken by the interaction term on the honeycomb lattice.

%from the mean-field decomposition $\alpha\beta\beta\beta\simeq -(i\alpha\beta)(i\langle\beta\beta\rangle)=-i\beta\beta\Delta_0$ and the fact that $\Delta_0\neq 0$ for the non-interacting Hamiltonian.

\section{Exact diagonalization and strong coupling limit}
\begin{figure}
\centering
\includegraphics[scale=0.4]{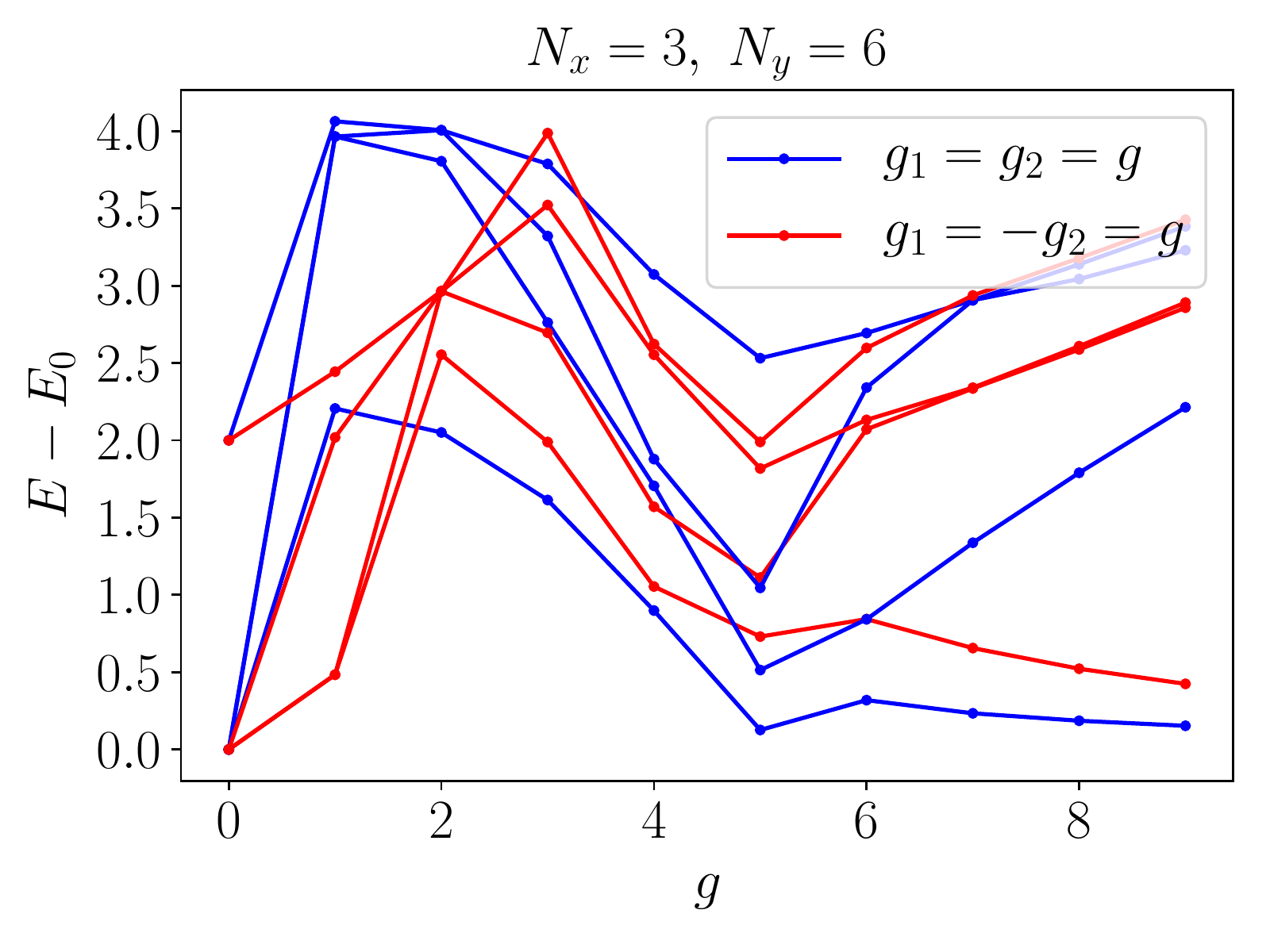}
\caption{Lowest many-body energy levels of $\cH$ computed using ED
  for different values of $g$.}\label{fig:energy}
\end{figure}

We perform exact numerical diagonalization of the full interacting
lattice model on clusters with up to  $N=32$ sites to ascertain
the validity of the MF results discussed above and to gain insights into
the strong coupling limit. Figures \ref{fig-5}(d)-\ref{fig-5}(f) show our ED results for
order parameters $\Delta_a$ compared to the results of the MF
analysis. At weak to intermediate coupling ($|g_1|,|g_2|\lesssim 2 t$)
we see that unbiased ED approach lends full support to our MF
results. At stronger coupling the two approaches begin to diverge
which suggests a breakdown of the MF theory in this limit. 
%Our limited exploration of the many-body energy spectra using ED along two lines $g=g_1=\pm g_2$ for larger $g$ suggest a transition near $g=5t$ to a phase with a doubly-degenerate ground state and a gapped excitation spectrum (Appendix E).

\begin{figure*}
\centering
\includegraphics[scale=0.3]{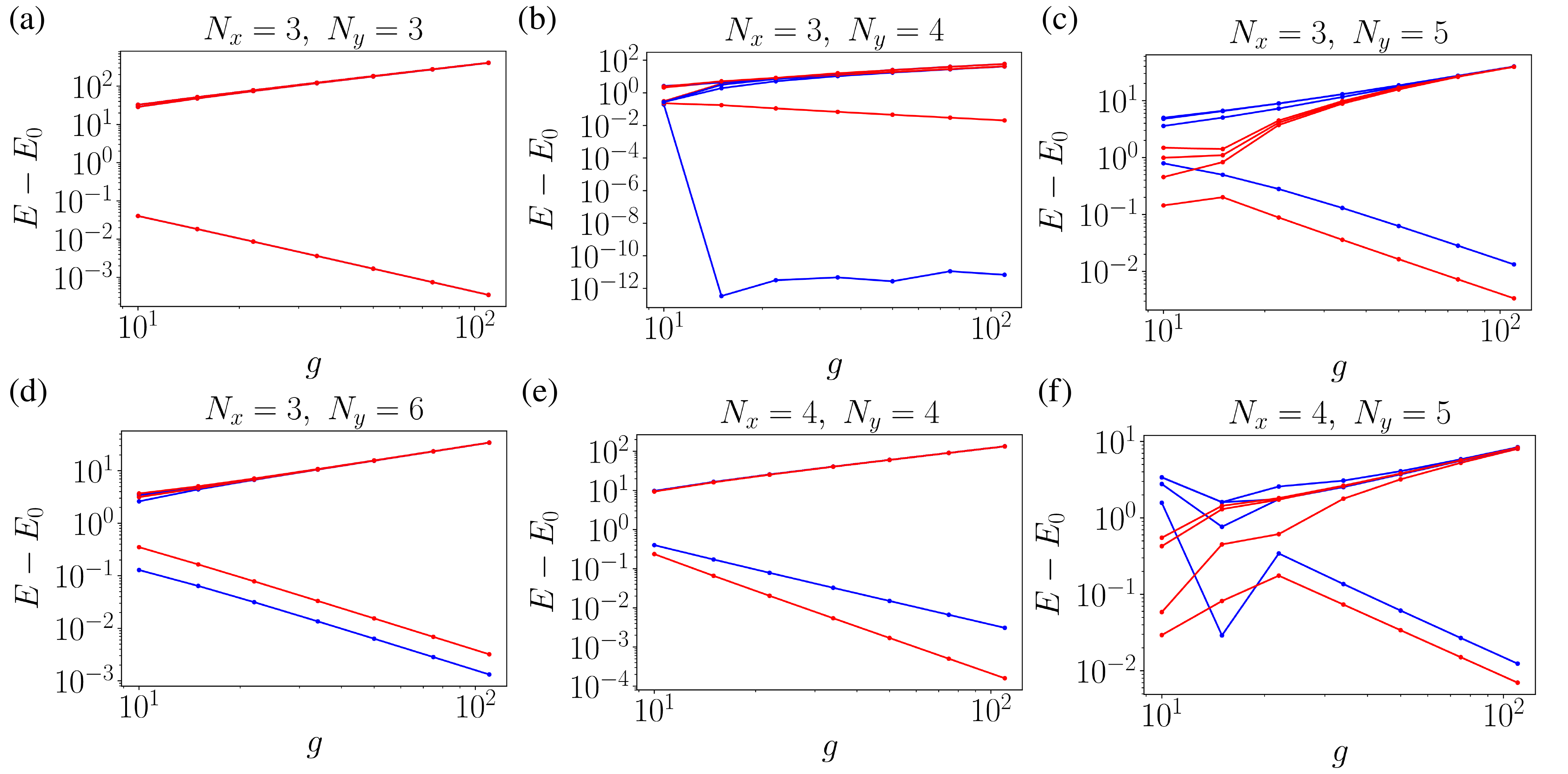}
\caption{Lowest many-body energy levels in the large $g$ limit, where $g_1=(-)g_2=g$ for the blue (red) curve as in Fig. \ref{fig:energy}.}
\label{fig:large-g}
\end{figure*}

We also calculate the lowest many-body energies using ED as a function of
$g_1=\pm g_2=g$; see Fig. \ref{fig:energy}. Although the detailed
behavior of the energy levels depends on the system geometry and size,
the results suggest that a phase transition occurs near $g\sim
5t$. Above the transition the pattern of energy levels shown in Fig. \ref{fig:large-g} is suggestive of a doubly
degenerate ground state and an excitation gap that grows linearly with $g$.

As a final topic, we briefly discuss
the physics of the model at large $g$. Analytical progress in this limit is
hampered by the fact that there is no obvious solution to the
problem when $t=0$. Since the $g_1$ and $g_2$ terms in   
$\cH_{\mathrm{int}}$ are seen to mutually commute the problem in this
limit separates into two commuting Hamiltonians that can be treated
independently. Nevertheless, these still remain difficult  problems with
no obvious solution.

\begin{figure*}
\includegraphics[width=14cm]{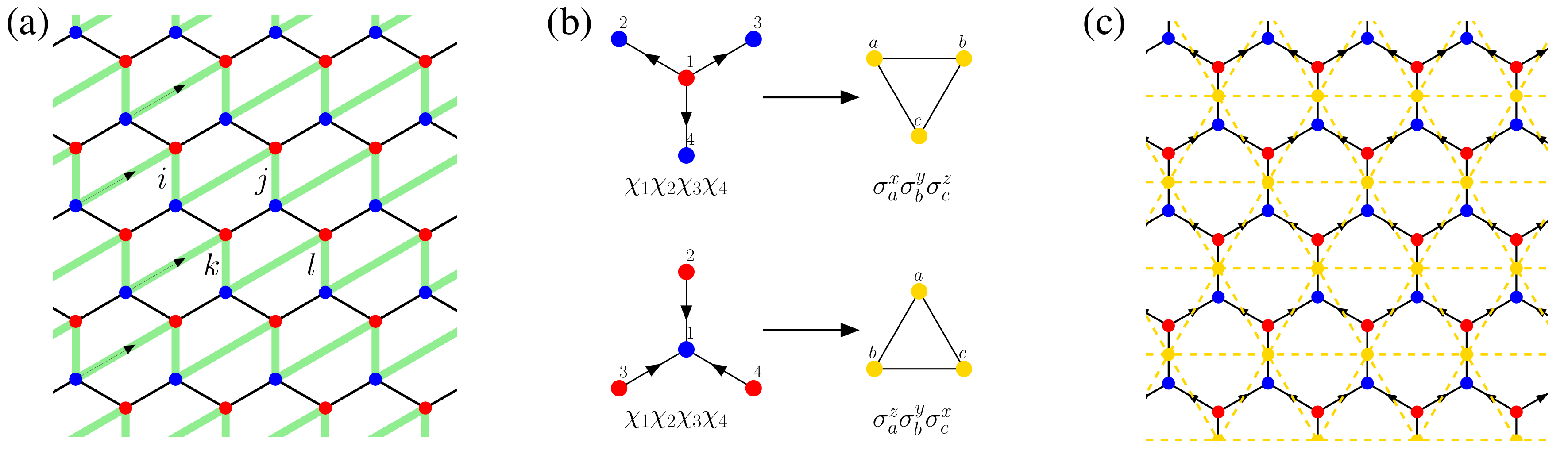}
\caption{Jordan-Wigner transformation. (a) The path used to define  the
  Jordan-Wigner transformation Eq.\ (\ref{JW}). (b) Each 
four-Majorana interaction term maps onto a three-spin interaction. (c) The
resulting triangular lattice spin model.}\label{fig:JW}
\end{figure*}

For $t=0$, it is possible to map $\cH_{\rm int}$ onto a local
spin-${1\over 2}$ model on a triangular lattice using the
Jordan-Wigner transformation. A set of Majorana operators ${\alpha_j,\beta_j}$ can be mapped onto a
set spin-${1\over 2}$ operators
$\bsig_j$ as
\begin{equation}\label{JW}
\alpha_i=\prod_{k<i}(-\sigma_k^z)\sigma_i^x,\ \ \
\beta_i=\prod_{k<i}(-\sigma_k^z)\sigma_i^y.
\end{equation}
For a generic local fermion Hamiltonian in dimension greater than 1,
however, the spin Hamiltonian might contain non-local
terms due to the products of $\sigma_k^z$ operators that appear in Eq.\ (\ref{JW}).
Fortunately, in the present case for $t=0$, the  Hamiltonian
remains strictly local if we choose the path shown in
Fig.\,\ref{fig:JW}(a) to order the sites. We thus get
\begin{equation}
\begin{split}
&\cH_{\Ydown}: \ \ g_1\alpha_k\beta_i\beta_j\beta_k=g_1\sigma_k^z\sigma_i^x\sigma_j^y,\\
&\cH_{\Yup}: \ \ g_2\beta_j\alpha_j\alpha_k\alpha_l=g_2\sigma_j^z\sigma_k^y\sigma_l^x,
\end{split}
\end{equation}
and the Hamiltonian is
\begin{equation}\label{hspin}
\cH_{\rm spin}=\left(g_1\sum_{\nabla}+g_2\sum_{\Delta}\right)\sigma_1^z\sigma_2^x\sigma_3^y,
\end{equation}
where the spins can be thought of as living on the midpoints of all vertical bonds of
the  original honeycomb lattice, and are arranged
on each triangle as indicated in Fig. \ref{fig:JW}(b). The spin system thus forms a
triangular lattice, Fig. \ref{fig:JW}(c). 

This is a highly frustrated Hamiltonian: while it is possible to
minimize the product of three spin operators on each of the triangles in
isolation it is not possible to do so for two triangles sharing a
single vertex. We thus conjecture that at strong coupling the MF state
discussed in the previous sections will give way to a highly entangled strong
coupling phase that can be viewed in the spin representation as a spin liquid. 
  The spin model (\ref{hspin}) shares some obvious similarities with
the celebrated Kitaev honeycomb lattice model
[\onlinecite{Kitaev2006}], but as far as we can tell it does not have an exact solution. 

ED calculations on small clusters at large $g$ indicate a featureless
ground state with
$\langle\alpha_1\beta_2\beta_3\beta_4\rangle=\langle\beta_1\alpha_2\alpha_3\alpha_4\rangle\simeq
\pm 1/2$ on each plaquette (the sign depends on whether $g_1=g_2$ or
$g_1=-g_2$). Two-fermion expectation values on first and second
neighbors are likewise featureless and in addition small compared to
unity. No obvious pattern of symmetry breaking is revealed by our
investigation. Collectively these results suggest a nontrivial, highly
entangled featureless state in the strong coupling limit which can be
possibly viewed as a spin liquid when represented through the
spin Hamiltonian  (\ref{hspin}). More work is clearly necessary to
determine the properties of this state.

\section{Conclusion}
Majorana-Hubbard model on the honeycomb lattice exhibits
interesting interaction-driven topological phases that occur already at
weak coupling. This is unlike other Majorana-Hubbard models previously
discussed in the literature [\onlinecite{Raghu2008,Rahmani2015b,Rahmani2015a,Cobanera2015,Fendley2018,Zhu2016,Affleck2017}].
The key distinction here is that the most local interaction term on the honeycomb lattice explicitly breaks the time-reversal symmetry ${\tilde\cT}$ which normally acts to protect the gapless nature of the excitation spectrum. 
 
The model may be realized at a proximitized surface of a 3D
 topological insulator [\onlinecite{Fu2008,Chiu2015}] if a vortex lattice with the
 honeycomb geometry can be stabilized. This could be achieved for
 instance by engineering such a surface with an array of pinning sites
 designed to bind vortices into the honeycomb lattice arrangement
 [\onlinecite{Baert1995,Harada1996}]. 

A related Majorana model on the honeycomb lattice with six-fermion
interactions was introduced in Ref.\ \onlinecite{Vijay2015} together
with a proposal for an experimental realization at a topological
insulator surface. In this setting our model becomes relevant when the
Majorana mode wavefunctions have large overlaps. The the hopping
parameter $t$ relative to the interaction parameter $g$ can be tuned
e.g.\ by shifting the chemical potential of the topological insulator as discussed in Ref. \onlinecite{Rahmani2015a}. Our
 predictions for interaction-driven topological phases can be tested by spectroscopic measurements using a scanning
 tunneling microscope, which is capable of locally distinguishing between
 gapped bulk and gapless edges of the system.

\emph{Acknowledgments.}-- We thank Ian Affleck, O\u{g}uzhan Can, \'{E}tienne Lantagne-Hurtubise and Tarun Tummuru for helpful discussions. The work described in this article was supported by NSERC and CIfAR.

\appendix 

\section{LATTICE MODEL SYMMETRIES}

We examine the symmetries of $\cH_0$ and $\cH_{\mathrm{int}}$ at the same
time. In both Hamiltonians we have discrete translational symmetry and
${2\pi\over 3}$ rotational symmetry. A general honeycomb lattice model
with these symmetries can in addition possess an inversion and two different reflection symmetries; see Fig. \ref{fig-2}(a). It can be easily
checked that the reflection symmetry with regard to the hexagon
diagonal, $\mathcal{R}_1$, is respected by the noninteracting model,
but broken by the interactions. Indeed, under $\mathcal{R}_1$,
$\cH(t,g_1,g_2)$ maps to $\cH(t,-g_1,-g_2)$.

Under inversion $\mathcal{P}$ sublattices interchange
$A\leftrightarrow B$ and therefore $\alpha_j\leftrightarrow\beta_j$. In
order to compensate for the minus sign in $\cH_0$, we introduce a
transformation $\mathcal{A}:\alpha\rightarrow-\alpha$, which amounts
to a Z$_2$ gauge transformation and thus does not change the
physics. One can check easily that $\mathcal{PA}$ is preserved if and
only if $g_1=g_2$ in the interacting case. Similar results hold for
the other  reflection $\mathcal{R}_2$, where
$\mathcal{R}_2\mathcal{A}$ is preserved if and only if $g_1=-g_2$. For
general $g_{1,2}$, $\mathcal{P}\mathcal{A}$ and
$\mathcal{R}_2\mathcal{A}$ map $\cH(t,g_1,g_2)$ to $\cH(t,g_2,g_1)$
and $\cH(t,-g_2,-g_1)$, respectively. Thus we are allowed to focus
only on the case $g_1\geq |g_2|$, and the behavior of the system in
the remaining regions of the phase diagram can be obtained from symmetry considerations. For example, the energy spectrum is the same for $\cH(t,g_1,g_2)$ and $\cH(t,-g_2,-g_1)$, while the Chern number acquires a minus sign.

Time reversal symmetry (TRS) is more subtle. Physically, we expect
Majorana modes to appear in the presence of the magnetic field; thus
physical TRS is broken from the very beginning. Nevertheless $\cH_0$
is invariant under antiunitary symmetry $\tilde{\cT}$: $(\alpha_j,\beta_j)\to
(\alpha_j,-\beta_j)$ and $i\to -i$ which acts, effectively, as a time
reversal with $\tilde{\cT}^2=+1$. The interaction term breaks this
symmetry as $\cH(t,g_1,g_2)$ maps to $\cH(t,-g_1,-g_2)$ under $\tilde{\cT}$. It follows
that a combined operation $\tilde{\cT}\mathcal{R}_1$ is a symmetry of
the full Hamiltonian for any $g_1$, $g_2$. The action of various
symmetries is summarized in Table \ref{tab:1}.

Vortices and antivortices remain the same under inversion (with an
immaterial minus sign), but map onto each other under
reflections. Thus we expect $\mathcal{P}$ to be relevant for a
lattice of (anti-)vortices, while $\mathcal{R}_{1,2}$ are respected by
a bipartite lattice of vortices  (antivortices) occupying sublattice A
(B). This motivates our exploration of the phase diagram for arbitrary
$g_{1,2}$ with special focus on two lines $g_2=\pm g_1$.
\begin{center}
\begin{table}
\begin{tabular}{ c||c|c|c } 
 \hline
 Symmetry & {translation} & \multicolumn{2}{c}{$2\pi/3$ rotation}  \\\hline
 Condition & \multicolumn{3}{c}{none} \\\hline
 Symmetry & $\mathcal{R}_1,\mathcal{T}$ & $\mathcal{P}\mathcal{A}$ & $\mathcal{R}_2\mathcal{A}$  \\\hline
 Condition & $g_1=g_2=0$ & $g_1=g_2$ & $g_1=-g_2$ \\
 \hline
\end{tabular}
\caption{The symmetries of the model, with definitions in the text.}\label{tab:1}
\end{table}
\end{center}

%------------------------------
\section{LOW-ENERGY FIELD THEORY}
The non-interacting Hamiltonian $\cH_0$ given in Eq.\ (1) of the main
text can be analyzed by introducing momentum-space Majorana operators
\begin{equation}\label{FT}
\begin{pmatrix}
\alpha_j \\ \beta_j
\end{pmatrix} 
=\sqrt{\frac{2}{N}}\sum_\bk \re^{i{\br}_j\cdot\bk}
\begin{pmatrix}
\alpha_\bk \\ \beta_\bk
\end{pmatrix},
\end{equation}
where $\br_j$ labels the unit cell. It is important to note that the
Fourier transform introduces a redundancy: because
\begin{equation}\label{FT1}
(\alpha_\bk^\dagger,\beta_\bk^\dagger)=(\alpha_{-\bk},\beta_{-\bk})
\end{equation}
the Fourier-space operators are no longer self-conjugate. One can deal
with the redundancy in two ways: (i) either view
$(\alpha_{\bk},\beta_{\bk})$ as independent across the entire BZ but
only consider positive energy eigenstates, or (ii) restrict $\bk$ to
one half of the BZ and consider all the states.

In the momentum space the Hamiltonian can be written as
$\cH_0=\sum'_\bk\Psi_\bk^{\dagger}h_0(\bk)\Psi_\bk$ where
$\Psi_\bk=(\alpha_{\bk},\beta_{\bk})^T$, the prime denotes summation
over half BZ and 
\begin{equation}\label{hk}
h_0(\bk)=
2t\begin{pmatrix}
0 & D_1(\bk)\\
D_1^{*}(\bk) & 0
\end{pmatrix},
\end{equation}
with $D_1(\bk)=i(1+\re^{i\bk\cdot\bd_1}+\re^{i\bk\cdot\bd_2})$. Here
$\bd_j$ are the primitive vectors of the Bravais lattice given by $\bd_{1,2}=(\mp\frac{\sqrt{3}}{2}a,\frac{3}{2}a)$.
%We use $\sum'$ to denote the sum over the whole unit cell and $\sum$ for half of it, which is the appropriate one for Majorana fermions. To be concrete, we choose the original unit cell as $[-\sqrt{3}/6,\sqrt{3}/6]\times[-1/3,1/3]$ and the ``half'' cell to be $[0,\sqrt{3}/6]\times[-1/3,1/3]$, both in unit of $2\pi/a$, see Fig.\,\ref{bz}. 
The diagonalization is straightforward and we have
\begin{equation}\label{E_0}
E_{\bk,\pm}=\pm 2t|1+\re^{i\bk\cdot\bd_1}+\re^{i\bk\cdot\bd_2}|.
\end{equation}
The energy spectrum is identical to that of the Dirac fermion model on
the honeycomb lattice familiar from graphene and exhibits nodal points
at $\pm\bK$ with $\mathbf{K}=(\frac{2\pi}{3\sqrt 3
  a},\frac{2\pi}{3a})$. Expansion of the Hamiltonian (\ref{hk}) near
$+\bK$, writing $\bk=\mathbf{K}+\mathbf{q}$ and assuming
$|\bq|$ small, gives a massless Dirac Hamiltonian 
\begin{equation}\label{hk1}
h_0(\bk)\simeq v
\begin{pmatrix}
0 & q_y+i q_x\\
q_y-iq_x & 0
\end{pmatrix},
\end{equation}
with velocity $v=3ta$ and the 
spectrum $E_\bq\simeq \pm v|\bq|$.

To derive the low-energy continuum theory we approximate the Majorana
fields by expanding close to the two nodal points, 
\begin{equation}
\begin{split}
\alpha(\br) &\simeq 2(e^{i\bK\cdot\br}\alpha_{+}(\br)+e^{-i\bK\cdot\br}\alpha_{-}(\br)),\\
\beta(\br)&\simeq 2(e^{i\bK\cdot\br}\beta_{+}(\br)+e^{-i\bK\cdot\br}\beta_{-}(\br)),
\end{split}
\end{equation}
where $(\alpha_\sigma,\beta_\sigma)$ with $\sigma=\pm$ are slowly
varying on the lattice scale and the normalization is chosen for later convenience. Substituting into the Hamiltonian  we
get
\begin{equation}
\begin{split}
\cH_0&\simeq
4it\int\dr \left[e^{i\bK\cdot\br}\alpha_{+}(\br)+e^{-i\bK\cdot\br}\alpha_{-}(\br)\right]\\
&\quad\times\bigl[e^{i\bK\cdot\br}\beta_{+}(\br)+e^{-i\bK\cdot\br}\beta_{-}(\br)\\
&\quad+e^{i\bK\cdot(\br+\bd_1)}\beta_{+}(\br+\bd_1)+e^{-i\bK\cdot(\br+\bd_1)}\beta_{-}(\br+\bd_1)\\
&\quad+e^{i\bK\cdot(\br+\bd_2)}\beta_{+}(\br+\bd_2)+e^{-i\bK\cdot(\br+\bd_2)}\beta_{-}(\br+\bd_2)\bigr].
\end{split}
\end{equation}
Now we expand the fields to leading order in $\bd_j$, e.g.\
$\beta_\sigma(\br+\bd_j)\simeq\beta_\sigma(\br)+\bd_j\cdot\nabla\beta_\sigma(\br)$,
and retain only the slowly-varying terms (i.e.\ those not containing
$e^{\pm i\bK\cdot\br}$ factors). We thus obtain the leading low-energy
free Hamiltonian
\begin{equation}
\cH_0\simeq-6ta\int\dr \big[\alpha_{-}(-\partial_x+i\partial_y)\beta_{+}
+\alpha_{+}(\partial_x+i\partial_y)\beta_{-}\big].
\end{equation}
Integrating by parts then leads to Eq.\ (3) of the main text.

It is also possible to express the kinetic term in the form of a Dirac Hamiltonian,
\begin{equation}
\cH_0\simeq-iv\int\dr \sum_{\sigma=\pm} \Psi_\sigma^\dagger\left(\sigma\tau^y\partial_x+\tau^x\partial_y\right)\Psi_\sigma,
\end{equation}
where $\tau^a$ are Pauli matrices, $\Psi_\sigma=(\alpha_\sigma,\beta_\sigma)^T$ and $\Psi_\sigma^\dagger=(\alpha_{\bar\sigma},\beta_{\bar\sigma})$. One could go one step further and write down the Lagrangian of the theory which shows explicitly the emergent low-energy Lorentz invariance, expected from a model defined on the honeycomb lattice. 

Analogous procedure can be applied to $\cH_{\rm int}$ and leads to the
low-energy expansion given in Eq.\ (4). It is to be  noted that unlike
the effective low-energy theory on the square lattice (where the
interaction  term contains no derivatives) here one derivative is
mandated because of the lattice structure of the interaction term. It
comprises either three $\alpha$ operators and one $\beta$ or vice
versa. It is easy to see that there is no way in this case to write a
non-derivative four-fermion term in the low-energy expansion. One can of
course have $\alpha_+\alpha_-\beta_+\beta_-$ but this corresponds to
a longer-range interaction term in the original lattice Hamiltonian, comprising
two A sites and two B sites of the honeycomb lattice, which will be
weaker on general grounds and we are therefore neglecting it here.

\section{DERIVATION OF THE MEAN-FIELD GAP EQUATIONS}
To begin we introduce the order parameters
\begin{equation}\begin{split}
&\Delta_0=i\langle\alpha_{i,1}\beta_{i,2}\rangle,\\    
&\Delta_1=i\langle\alpha_{i,1}\alpha_{i,2}\rangle,\\
&\Delta_2=i\langle\beta_{i,1}\beta_{i,2}\rangle,
\end{split}
\end{equation}
assuming translational invariance and signs
 illustrated in Fig.\,\ref{fig-2}(c). 
Using these we can write the mean field ground state energy as
\begin{equation}
E_{\MF}=\langle\Psi_\MF|\cH_{\MF}|\Psi_\MF\rangle=3N\sum_{a=0}^{2}\tau_a\Delta_a,
\end{equation}
and it holds that
\begin{equation}
\Delta_a=\frac{1}{3N}\frac{\partial E_{\MF}}{\partial \tau_a},
\end{equation}
by the Hellmann-Feynman theorem. Using Eq.\, (\ref{E_pm}), we can
explicitly perform the derivatives and get a set of equations 
\begin{equation}\label{Delta}
\begin{split}
\Delta_0&=\frac{2\tau_0}{3N}\Big(\sum_{\bk:+}\frac{|D_1|^2}{\epsilon_{\bk}}
-\sum_{\bk:-}\frac{|D_1|^2}{\epsilon_{\bk}}\Big),\\
\Delta_1&=\frac{2}{3N}\Big(\sum_{\bk:+}\Big(D_2
+\frac{(\tau_1+\tau_2)D_2^2}{\epsilon_{\bk}}\Big)\\
&+\sum_{\bk:-}\Big(D_2-\frac{(\tau_1+\tau_2)D_2^2}{\epsilon_{\bk}}\Big)-\sum_{\bk}D_2\Big),\\
\Delta_2&=\frac{2}{3N}\Big(\sum_{\bk:+}\Big(-D_2+\frac{(\tau_1+\tau_2)D_2^2}{\epsilon_{\bk}}\Big)\\
&+\sum_{\bk:-}\Big(-D_2-\frac{(\tau_1+\tau_2)D_2^2}{\epsilon_{\bk}}\Big)+\sum_{\bk}D_2\Big),
\end{split}
\end{equation}
where $\sum_{\bk:\pm}$ denotes summation over the occupied (i.e.\
negative-energy) states in the upper (+)
or lower ($-$) band.

The MF energy of the full  interacting model can also be easily
written down in terms of $\{\Delta_a\}$,
\begin{equation}
\langle \cH\rangle=\langle\Psi_\MF|\cH_0+\cH_{\mathrm{int}}|\Psi_\MF\rangle=3N\Delta_0(t+g_1\Delta_2+g_2\Delta_1).
\end{equation}
Now we can get the gap equations by minimizing the energy with respect
to variational parameters $\{\tau_a\}$,
\begin{equation}
\frac{\partial \langle\cH\rangle}{\partial \tau_a}=0,
\end{equation}
or more explicitly
\begin{equation}
\frac{\partial \Delta_0}{\partial \tau_a}(t+g_1\Delta_2+g_2\Delta_1)
+\Delta_0\Big(g_1\frac{\partial \Delta_2}{\partial \tau_a}
+g_2\frac{\partial \Delta_1}{\partial \tau_a}\Big)=0.
\end{equation}
We also note a corollary of the Hellmann-Feynman theorem
\begin{equation}
\sum_{a=0}^2\tau_a\frac{\partial \Delta_a}{\partial \tau_a}=0.
\end{equation}
It is easy to check that the last two equations are solved by
variational parameters $\{\tau_a\}$ given by Eqs.\ (12) in the main
text and order parameters  $\{\Delta_a\}$ given by Eq.\ (\ref{Delta}).

\eject
\vfill
\bibliography{MH}

\newpage

%\end{widetext}
\end{document}